\documentclass[12pt,preprint]{aastex}

\newcommand{\hii}{H\,II}
\newcommand{\w}{\,$\lambda$\,}
\newcommand{\ww}{\,$\lambda\lambda$\,}

\newcommand{\msun}{$M_\odot$}
\newcommand{\ebv}{$E(B-V)$}

\newcommand{\II}{\,{\sc ii}}
\newcommand{\III}{\,{\sc iii}}
\newcommand{\IV}{\,{\sc iv}}            
\newcommand{\V}{\,{\sc v}} 

\shorttitle{STIS UV spectroscopy of early B supergiants in M31}
\shortauthors{Bresolin et al.}

\begin{document}

\title{STIS UV spectroscopy of early B supergiants in M31\footnotemark}
\footnotetext[1]{Based on observations with the NASA/ESA Hubble Space
Telescope obtained at the Space Telescope Science Institute, which is
operated by the Association of Universities for Research in Astronomy,
Incorporated, under NASA contract NAS5-26555.}

\author{Fabio Bresolin}
\affil{Institute for Astronomy, 2680 Woodlawn Drive, Honolulu HI
96822}
\email{bresolin@ifa.hawaii.edu}

\author{Rolf-Peter Kudritzki}
\affil{Institute for Astronomy, 2680 Woodlawn Drive, Honolulu HI
96822}
\email{kud@ifa.hawaii.edu}

\author{Daniel J. Lennon}
\affil{Isaac Newton Group, Apartado 321, 38700 Santa Cruz de La Palma, Canary 
Islands, Spain}
\email{djl@ing.iac.es}

\author{Stephen J. Smartt}
\affil{Institute of Astronomy, University of Cambridge, Madingley
Road, Cambridge CB3 0HA}
\email{sjs@ast.cam.ac.uk}

\author{Artemio Herrero}
\affil{Instituto de Astrof\'{\i}sica de Canarias, E-38200 La Laguna,
Tenerife, Spain}
\affil{Departamento de Astrof\'{\i}sica, Universidad de La Laguna,
Avda. Astrof\'{\i}sico Francisco S\'{a}nchez, s/n, E-38071 La Laguna,
Spain}
\email{ahd@ll.iac.es}

\author{Miguel A. Urbaneja}
\affil{Instituto de Astrof\'{\i}sica de Canarias, E-38200 La Laguna,
Tenerife, Spain}
\email{maup@ll.iac.es}

\and 

\author{Joachim Puls}
\affil{Universit\"{a}ts-Sternwarte M\"{u}nchen, Scheinerstrasse 1,
81679, M\"{u}nchen, Germany}
\email{uh101aw@usm.uni-muenchen.de}

\begin{abstract}
We analyze STIS spectra in the 1150-1700~\AA\/ wavelength range
obtained for six early B supergiants in the neighboring galaxy
M31. Because of their likely high (nearly solar) abundance, these
stars were originally chosen to be directly comparable to their
Galactic counterparts, and represent a much-needed addition to our
current sample of B-type supergiants, in our efforts to study the
dependence of the Wind Momentum-Luminosity Relationship on spectral
type and metallicity.  As a first step to determine wind momenta we
fit the P-Cygni profiles of the resonance lines of N\V, Si\IV\/ and
C\IV\/ with standard methods, and derive terminal velocities for all
of the STIS targets. From these lines we also derive ionic stellar
wind column densities.  Our results are compared with those obtained
previously in Galactic supergiants, and confirm earlier claims of
`normal' wind line intensities and terminal velocities in M31.  For
half of the sample we find evidence for an enhanced maximum turbulent
velocity when compared to Galactic counterparts.

%The
%mean turbulent velocity derived from the line fits, approximately 23\%
%of the terminal velocity, however, is considerably larger than for
%Galactic counterparts, but in agreement with our recent findings in
%early B-type supergiants in M33.

\end{abstract}


%===================================================================================

\section{Introduction}

The analysis of mass-loss and stellar winds in early-type supergiants,
important {\em per se} because of the physical insight it provides
about the atmospheres of massive stars, has gained momentum in recent
years with the realization that it can also provide the basis for the
determination of stellar distances, through the dependence of the wind
momentum on luminosity. With the work of \citet{puls96} and
\citet{kudritzki99}, the Wind Momentum-Luminosity Relationship (WLR)
has been established for OBA supergiants in our own Galaxy and in the
Magellanic Clouds. The latter paper showed the impact of the differing
stellar atmospheric parameters between spectral types on the
parameterization of the WLR. The empirical verification of the
predicted dependence of the global properties of stellar winds, and
consequently of the WLR, on metallicity has however just begun,
because of the considerable observational efforts required
(\citealt{mccarthy95,mccarthy97}). It is nevertheless reassuring that
preliminary results on the massive stellar winds in galaxies even
beyond the Local Group agree with the present calibration of the WLR
(\citealt{bresolin01,bresolin02}).

As a next step in the understanding and calibration of the WLR we are
carrying out several programs, including the study of OB supergiants
of known distances in the Galaxy (\citealt{herrero01}), M33
(\citealt{urbaneja02}) and M31 (this work). This allows us to
investigate stellar winds at different metallicities and spectral
types, by removing the uncertainty in the distances which afflicts
most of the Galactic work (\citealt{kudritzki99}).

In this paper we present our first results on early B (B0-B3)
supergiants in the neighboring galaxy M31, obtained from Space
Telescope Imaging Spectrograph (STIS) ultraviolet spectra. Given the
nearly solar abundance of the observed fields, as inferred from \hii\/
region studies (\citealt{blair82}), these stars were originally chosen
to be directly comparable to their Galactic counterparts.  A
comparison of the WLR obtained for two A-type supergiants in M31 with
similar objects in the Galaxy has been presented in the work of
\citet{mccarthy97}, and by \citet{kudritzki99}. However, for the
purpose of calibrating the WLR, early B supergiants are currently
under-represented in the available samples of stars with accurate
distances. Moreover, as shown by
\citet{kudritzki99}, changes in the ionization stages of the metals
driving the wind around the late O and early B types modify the WLR
considerably, and must be thoroughly tracked by gathering more data
for stars in this spectral range.  For A-type supergiants the wind
parameters, including the mass-loss rate and the terminal velocity
$v_\infty$, can be derived from fits of the H$\alpha$ line to
hydrodynamical models of the expanding atmosphere. However, for OB
supergiants, profile fits to the UV P-Cygni resonance lines are
required to obtain accurate terminal velocities (\citealt{kud00}),
hence the necessity for HST spectroscopy.

Individual OB supergiants in M31 have been observed previously in the
vacuum ultraviolet with the IUE satellite (\citealt{bianchi91}, and
references therein) and with the Faint Object Spectrograph (FOS) on
the Hubble Space Telescope (HST; \citealt{hutchings92},
\citealt{bianchi94}). 
Some of this early work suffered from low spectral resolution and
sensitivity, and suggested that the stellar winds of OB supergiants in
M31 could be one order of magnitude weaker than in the Galaxy. This
result was challenged by \citet{herrero94}, who concluded from a study
of H$\alpha$ spectra in M31 AB supergiants that their winds should be
comparable to those of similar Galactic stars. A reanalysis of FOS
data by \citet{bianchi94} and \citet{haser95b} established that
mass-loss rates in the studied M31 OB supergiants are comparable to
those found in some of their Galactic counterparts.

The targets for the present investigation are six early B supergiants
chosen among the brightest from the surveys of massive stars in M31 by
\citet[CCD {\em UBV} imaging]{massey86}, \citet{humphreys90} and \citet[optical
spectroscopy]{massey95}.  The nomenclature adopted here for the parent
OB associations (see Table~1) is the one introduced by
\citet{vandenbergh64}. One of the supergiants lies in Baade's Field IV
(\citealt{baade63}), within the boundaries of van den Bergh's
association OB\,184, one of the outermost associations in the disk of
M31, $96\arcmin$ (22~Kpc) southwest of the nucleus. This star is indicated as
IV-B59 in Table~1, and is one of the three early-type supergiants
included in the study of
\citet{humphreys79}. Of the remaining stars, two are found in
OB\,78 (=\,NGC~206), and one in OB\,48, both among the richest OB
associations in M31, located in the most actively star forming annular
region, between 9 and 15~Kpc, identified by
\citet{vandenbergh64}. The last two objects are part of OB\,8 and
OB\,10, at the smallest galactocentric distance ($\sim$\,6~Kpc) in the
sample considered here.

High resolution optical spectra of all the STIS targets have been
obtained by us with the William Herschel Telescope (WHT) on La Palma
and with the Keck~I telescope.  The analysis of the WHT spectra are
presented in a companion paper (\citealt{trundle02}), deriving stellar
parameters and photospheric abundances. A future analysis of the
complete data set will be achieved in a unified, consistent approach
to further yield mass-loss rates and a full test of the WLR for the
M31 galaxy. We also note that previous optical spectroscopic work on
individual B- and A-type M31 supergiants has been carried out by
\citet[the star OB\,78-478 is in common with our present
sample]{herrero94}, \citet{mccarthy97},
\citet{venn00} and \citet[star OB\,10-64 in common with current work]{smartt01}.

In this paper we concentrate on the measurement of the terminal
velocities from the analysis of the main resonance lines in the UV
spectrum. We describe the STIS observations of the six B supergiants
in M31 in \S~2, and the method used to derive the terminal velocities
in \S~3. We comment on the individual stars in
\S~4, and we briefly discuss our results in \S~5.

\section{Observations}
The data were collected with STIS aboard HST
(\citealt{woodgate98}, \citealt{kimble98}), using the 0.2~arcsec-wide
slit, combined with the G140L grating and the FUV MAMA detector. This
setup delivered spectra in the 1150-1700~\AA\/ wavelength range, at
$R\sim 1000$ resolution.  Total exposure times and observing dates are
summarized in Table~1, together with the main observational data for
the six M31 supergiants. Magnitudes are from CCD photometry
(\citealt{massey95}), except in the case of IV-B59 (photoelectric
measurement by \citealt{humphreys79}). The spectral types, initially
drawn from \citet[=\,M95 in Table~1]{massey95}, and from
\citet{humphreys79} for IV-B59, have been revised from the new optical
spectra obtained at the WHT (\citealt[=\,T02]{trundle02}). For the
following analysis of the UV spectra we adopted the radial velocities
determined from our optical spectra, with corresponding uncertainties
of 10-20 km\,s$^{-1}$.  WFPC2 images of some of our fields reveal that
OB\,78-159 is a composite object, consisting of at least three
stars. Multi-color WFPC2 images for OB\,48-358 and OB\,78-478 show
them to be single objects at the HST resolution, while the STIS
acquisition images for the remaining three targets also indicate they
are single.

The pipeline-processed extracted spectra were averaged and rectified,
using a low-order polynomial to fit regions of the continuum
relatively free of stellar features, as indicated by stellar
atmosphere models (e.g.,~\citealt{leitherer01}). Although the procedure
is made uncertain by heavy blocking from metal lines (mostly iron),
especially redwards of C\IV\w1550, and by interstellar Ly\,$\alpha$
absorption, the following results pertaining to the analysis of the
resonance wind lines remain virtually unaffected. 

\begin{figure}
\plotone{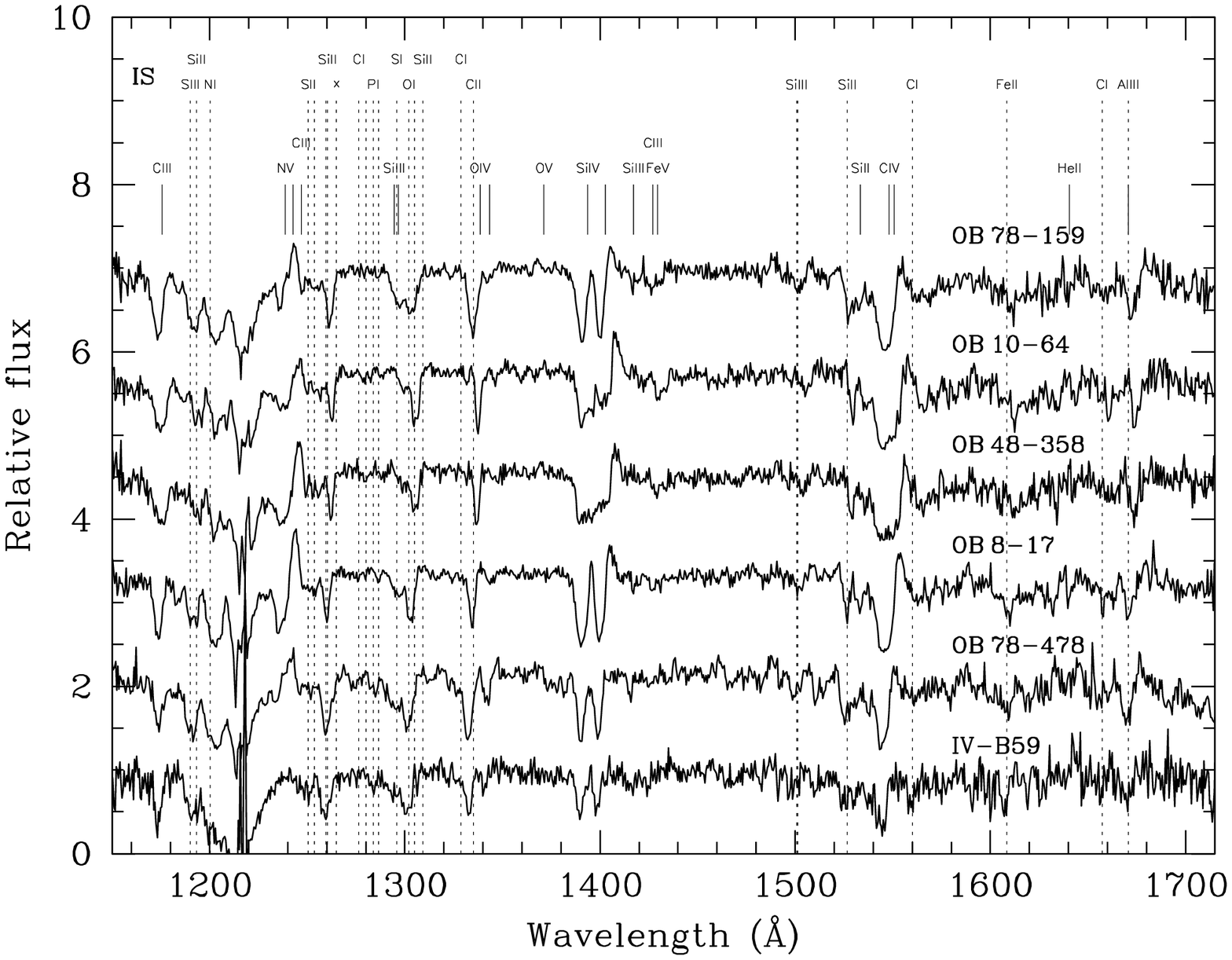}
\caption{The rectified STIS spectra of the B supergiants observed in
M31. A relative flux offset is applied for clarity. Identification for
the main interstellar (dashed vertical lines) and stellar (vertical
segments) features is provided.\label{spectra}}
\end{figure}

Rectified spectra are displayed in Fig.~\ref{spectra}. The wavelength
region covered includes some of the most important diagnostics of
stellar winds, i.e., the doublets N\V\ww1239,1243 (hereafter
abbreviated as N\V\w1240), Si\IV\ww1394,1403 (Si\IV\w1400) and
C\IV\ww1548,1551 (C\IV\w1550). The P-Cygni profiles observed in all
these lines are consistent with the expected profiles for early B-type
supergiants (\citealt{walborn87}), with a decreasing wind effect as
one goes from B0 to B2 supergiants. N\V\/ is virtually absent in the
later type star, IV-B59.  Several photospheric lines can be
identified, including C\III\w1176, O\IV\w1339,1343 and
He\II\w1640. The displacement towards short wavelengths of the
C\III\w1176 blend observed for the earlier type stars (OB\,78-159,
OB\,10-64 and OB\,48-358) is consistent with the wind effect seen in
O4-B0 supergiants, and which quickly declines after the B0.5\,Ia class
(\citealt{snow76}).

The presence of several absorption lines due to Fe\V\/ and Fe\IV\/ is
responsible for the continuum depression at wavelengths longer than
1400~\AA. Moreover, prominent interstellar lines are visible throughout the
spectra, originating from singly ionized Si (e.g.,~Si\II\w1193, \w1260 and
\w1527), and additional metals, as identified in Fig.~\ref{spectra}.
For the stars with a higher radial velocity ($v_{rad}\simeq
500$~km\,s$^{-1}$) these features are generally broader, because of
the resulting larger separation in wavelength between the M31 and the
Galactic contribution.

\section{Analysis of the resonance lines}

For the analysis of the resonance lines and the determination of the
terminal velocities we used the method described by
\citet[see also \citealt{haser95b}]{haser95}. This is based on the
SEI-method developed by \citet[see also \citealt{hamann81}]{lamers87}
for the calculation of line profiles in stellar winds, and which has
been adopted in subsequent investigations of the UV line profiles of
massive stars (\citealt{groenewegen89}, \citealt{lamers99},
\citealt{herrero01}).  Here we briefly describe the formalism adopted
in this work, for further details the reader is referred to the papers
cited above. The wind flow is described by a $\beta$-law:
$$w(x)=(1-b/x)^\beta
\hspace{1cm}b=(1-w^{1/\beta}_{min})$$

where $x=r/R_*$ is the radial coordinate normalized to the stellar
radius, $w(x)=v(x)/v_\infty$ is the flow velocity normalized to the
terminal velocity, and $w_{min}=w(x=1)$ is considered fixed at 0.01.
Following \citet{haser95}, a wind `turbulent' velocity increasing with
radius is assumed: $$v_{turb}=a_tv(r)+b_t\hspace{0.5cm}
a_t=\frac{v_{ta}-v_{ti}}{1-v_{min}}
\hspace{0.5cm}b_t=v_{ta}-a_{t},$$

with $v_{ti}$ and $v_{ta}$ the minimum and maximum turbulent
velocities, reached at $w=w_{min}$ and $w=w_{max}=1$, respectively.

\begin{figure}
\plotone{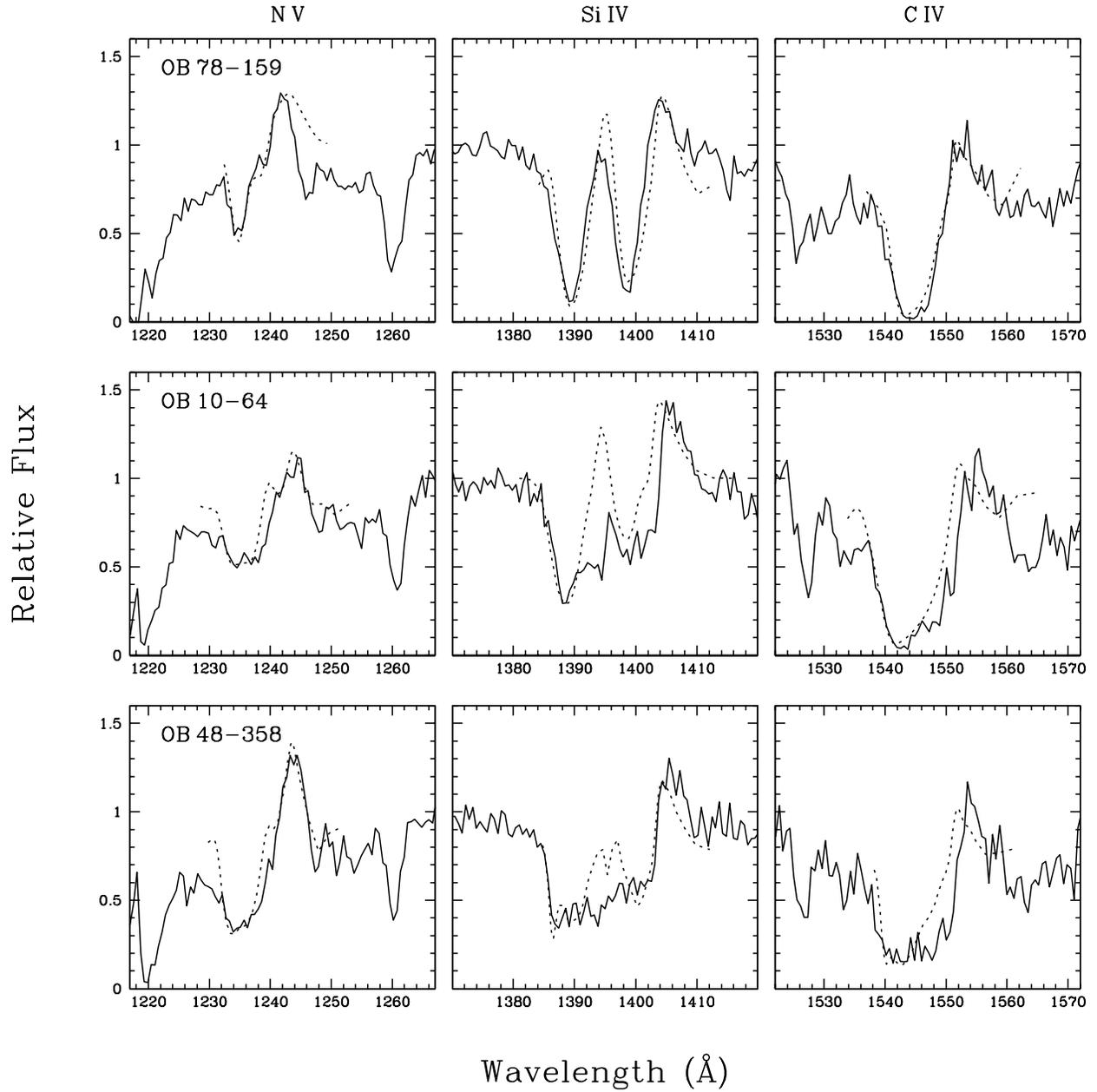}
\caption{Profile fits to the N\V\w1240, Si\IV\w1400 and C\IV\w1550
lines are shown as dotted lines superimposed on the observed
normalized spectra, for the stars OB\,78-159, OB\,10-64 and
OB\,48-358.\label{fit1}}
\end{figure}

\begin{figure}
\plotone{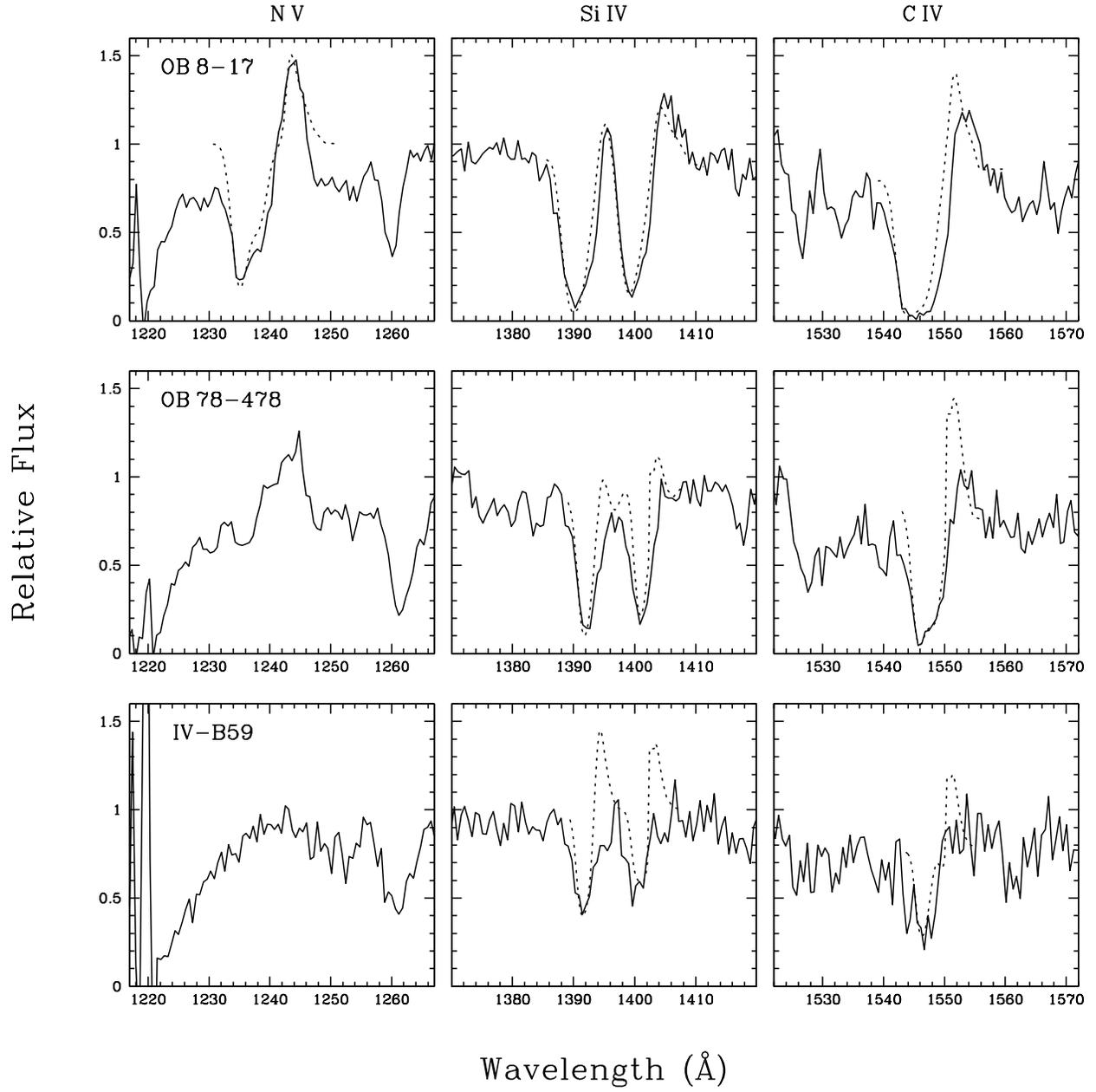}
\caption{As Fig.~\ref{fit1}, for stars OB\,8-17, OB\,78-478 and
IV-B59.\label{fit2}}
\end{figure}

To account for the contamination of the observed line profiles by
photosperic lines we used IUE spectra of B-type main sequence stars
having weak winds and small projected rotational velocities. The
calculated line profiles, once their resolution was degraded to match
the resolution of the STIS spectra, were then directly comparable to
the observed ones. Our adopted best fits for the N\V, Si\IV\/ and
C\IV\/ doublets are displayed, along with the observed line profiles,
in Fig.~\ref{fit1} and \ref{fit2}.  Even though the quality of the fit
to the emission peaks, as well as to the central part of the line
profiles, depends on the chosen photospheric templates, the derived
terminal velocities are quite insensitive to this choice.  Some
additional caution had to be taken when fitting the profiles, because
of the contamination from instrumentally broadened interstellar lines
and because of the depression of the pseudocontinuum created by strong
metal line blocking on the blue side of C\IV\w1550. Stellar rotation
is not expected to play a major role in our results, since
typical rotational velocities for early B supergiants are $\sim$50$-$80
km\,s$^{-1}$.

There is in general a good agreement in the terminal velocities
derived from the different resonance lines. Their mean values are
summarized in Table~\ref{results} (column~8, with errors given in
brackets), together with the range in maximum turbulent velocity found
to be acceptable for the fits (column~9), and the exponent $\beta$ of
the wind parameterization (column 10). The uncertainty in the terminal
velocities is of order 50 to 100 km\,s$^{-1}$, as estimated from model
fits obtained varying $v_{ta}$ and $v_\infty$ in combination.
Concerning the maximum turbulent velocity, for half of the sample its
largest value in the estimated range of validity is comparable to the
spectral resolution of the data (approximately 250 km\,s$^{-1}$), so
that secure inferences about the $v_{ta}/v_\infty$ ratio cannot be
drawn.  We conclude that these data are not inconsistent with previous
studies of Galactic supergiants, for which $v_{ta}/v_\infty=0.08-0.15$
(\citealt{groenewegen89}, \citealt{haser95}, \citealt{herrero01}). For
the remaining objects (OB\,78-159, OB\,10-64 and OB\,8-17), on the
other hand, we find some evidence for an enhanced $v_{ta}$, in
particular large turbulent velocities
($v_{ta}/v_\infty\simeq0.20-0.30$, with $v_{ta}$ in excess of the
spectral resolution) seem to be required in order to explain both the
blue slope and the width of the resonance line profiles (broader than
those found in spectra of Galactic supergiants observed with IUE and
degraded to STIS resolution).  This result is in agreement with a
similar finding in a sample of early B supergiants in M33 (mean
$v_{ta}/v_\infty=0.25$) studied by \citet{urbaneja02}.

Column densities (in cm$^{-2}$) for the three ions reported in
Table~\ref{results} (columns 11-13) were calculated as in \citet[see
also \citealt{herrero01}]{haser95}, from the integration of the line
strengths between $w=0.2$ and $w=1.0$. In Table~2, we have adopted a
distance to M31 of 783 Kpc (\citealt{holland98}) for deriving radial
galactocentric distances $R$ (column 2) and absolute magnitudes $M_V$
(column 5).

\subsection{HI column densities}

\begin{figure}
\plotone{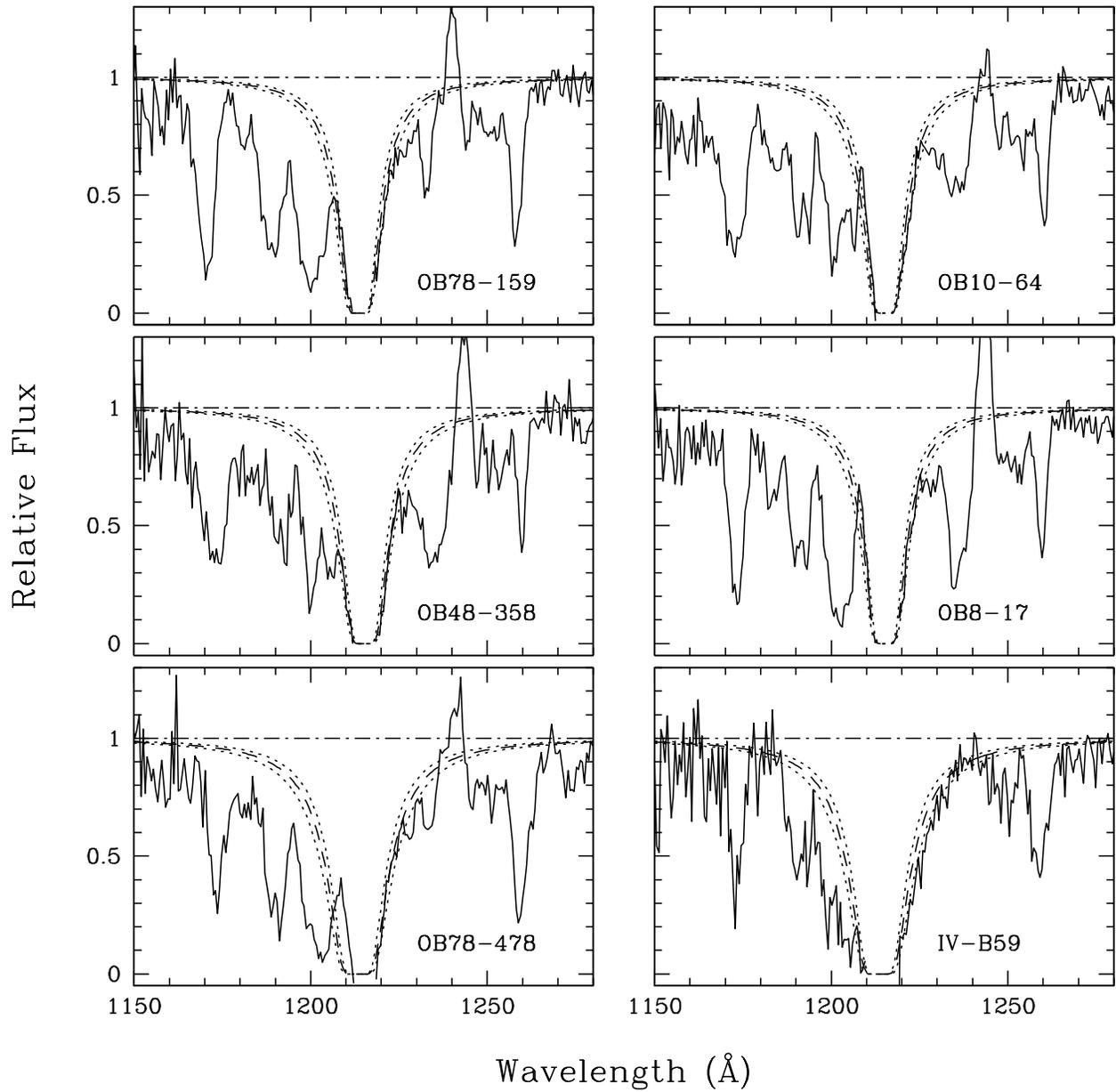}
\caption{Fits to the interstellar Ly\,$\alpha$
absorption, assuming a pure damping profile, for the adopted neutral
hydrogen column density (dashed line), and varying the column density
by $\pm0.1$ dex (dotted lines).\label{nh}}
\end{figure}

The column density of neutral hydrogen atoms along the line of sight,
$N(H\,I)$, was estimated from the observed interstellar medium
Ly\,$\alpha$ absorption, assuming a pure damping profile for this line
(\citealt{bohlin75}). The profiles for the adopted column densities
are illustrated in Fig.~\ref{nh} (dashed lines), together with the
profiles obtained by varying $N(H\,I)$ by $\pm0.1$ dex (dotted lines),
our estimated uncertainty from the comparison of damping profiles
having different column densities with the observed ones.  In the fits
more weight was given to the red part of the line wings, since the
blue part is more heavily affected by absorption features. A look at
Fig.~\ref{nh} shows that for the two stars for which we derive the
largest $N(H\,I)$ (OB\,78-478 and IV-B59) the fits are also the most
uncertain.  We have verified that the stellar contribution to the
Ly\,$\alpha$ line wings is minimal, and does not affect the estimated
$N(H\,I)$ in a significant way.

The resulting total column densities are given in column 6 of
Table~\ref{results}, while column 4 contains the color excess \ebv,
taken as the average for a given OB association, from \citet{massey95}
and \citet{humphreys79}. \ebv\/ values for the single stars are also
given by these authors, and are within 0.06 mag (of the same order as
the typical uncertainty on the measured colors) of the parent
association average, which we prefer in this case in order to avoid
errors from possible spectral type misclassifications.  To estimate
the gas-to-dust ratio as derived from these measurements, we first
removed the Galactic foreground contribution, taking the color excess
in the direction of M31 from
\citet{schlegel98}, $E(B-V)_{Gal}=0.06$ mag, and the average gas-to-dust
ratio from \citet{savage79}, [$N(HI)/E(B-V)]_{Gal}=5.0\times 10^{21}$
cm$^{-2}$~mag$^{-1}$. The resulting $N(HI)/E(B-V)_0$, intrinsic
to M31, is given in column 7, with corresponding uncertainties in
brackets, assuming a 0.04 mag error in the color index.

The weighted mean ratio, $N(HI)/E(B-V)_0=5.1\,(\pm1.3)\times 10^{21}$
cm$^{-2}$~mag$^{-1}$, compares well with the Galactic one, and with
previous determinations for M31.  Using globular clusters as tracers
of reddening, \cite{bajaja77} found a radial dependence of the M31
gas-to-dust ratio, averaging $3.3\times 10^{21}$ cm$^{-2}$~mag$^{-1}$,
while \citet{vandenbergh75} found a value similar to the Galactic one
(see also \citealt{braun92}, who used the Balmer decrement of ionized
gas as a measure of reddening). Higher values have been recently
reported by \citet{cuillandre01}, ranging between
$N(HI)/E(B-V)_0=1.1\times 10^{22}$ cm$^{-2}$~mag$^{-1}$ (using
background galaxies seen through the disk of M31) and $1.45\times
10^{22}$ cm$^{-2}$~mag$^{-1}$ (from M31 stars) in an outer field of
M31.

%We note that our neutral hydrogen column densities have not been
%corrected for the HI warp observed in M31 (\citealt{brinks84}).

\section{Comments on individual supergiants}

\begin{figure}
\plotone{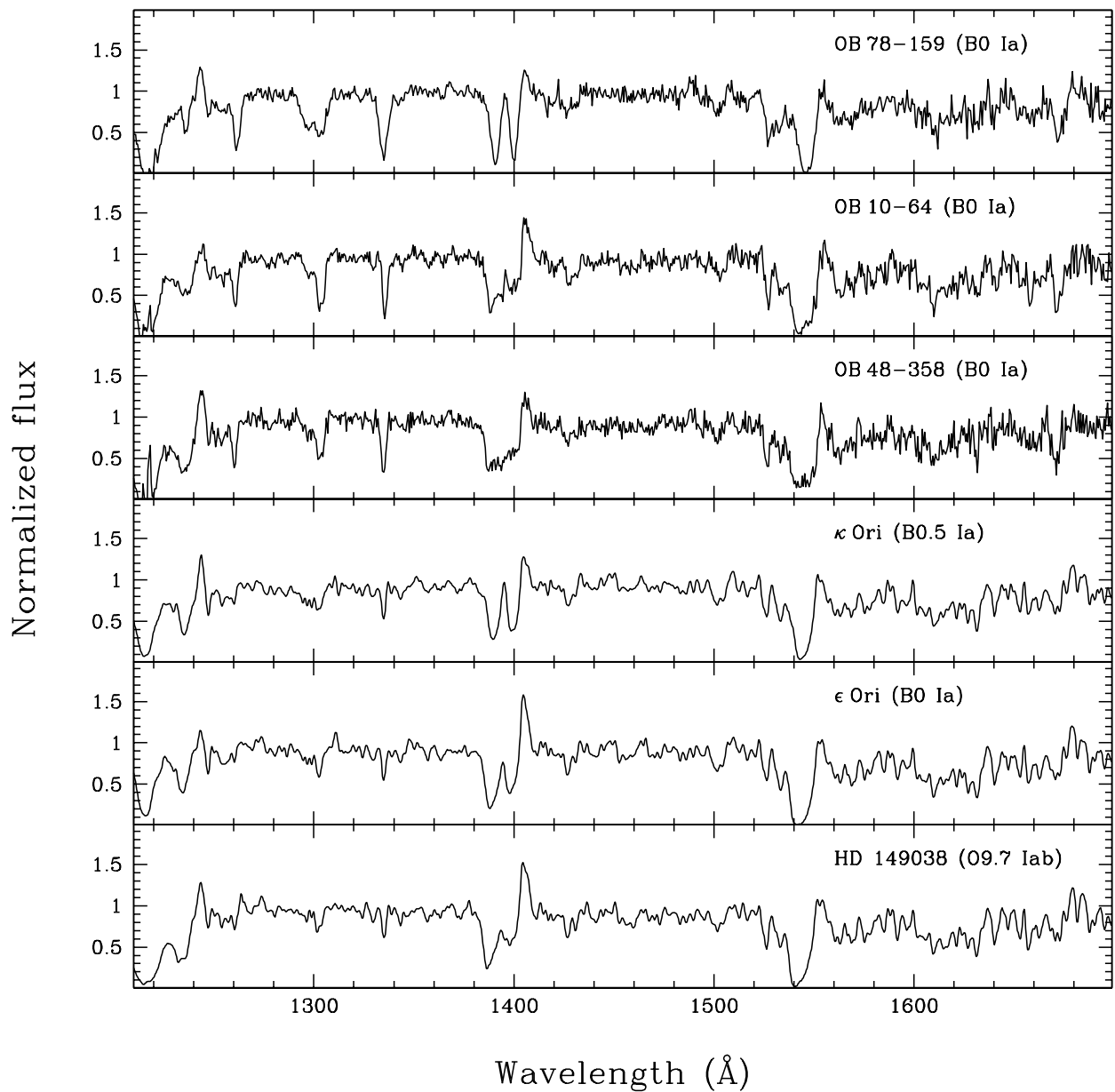}
\caption{The STIS spectra of the three B0\,Ia stars in the M31 sample 
(upper three panels) are compared to IUE spectra of Galactic stars of
similar spectral type, degraded to the resolution of the STIS spectra:
$\kappa$~Ori (B0.5\,Ia), $\epsilon$~Ori
(B0\,Ia) and HD~149038 (O9.7\,Iab) (lower three panels).\label{comp1}}
\end{figure}

\begin{figure}
\plotone{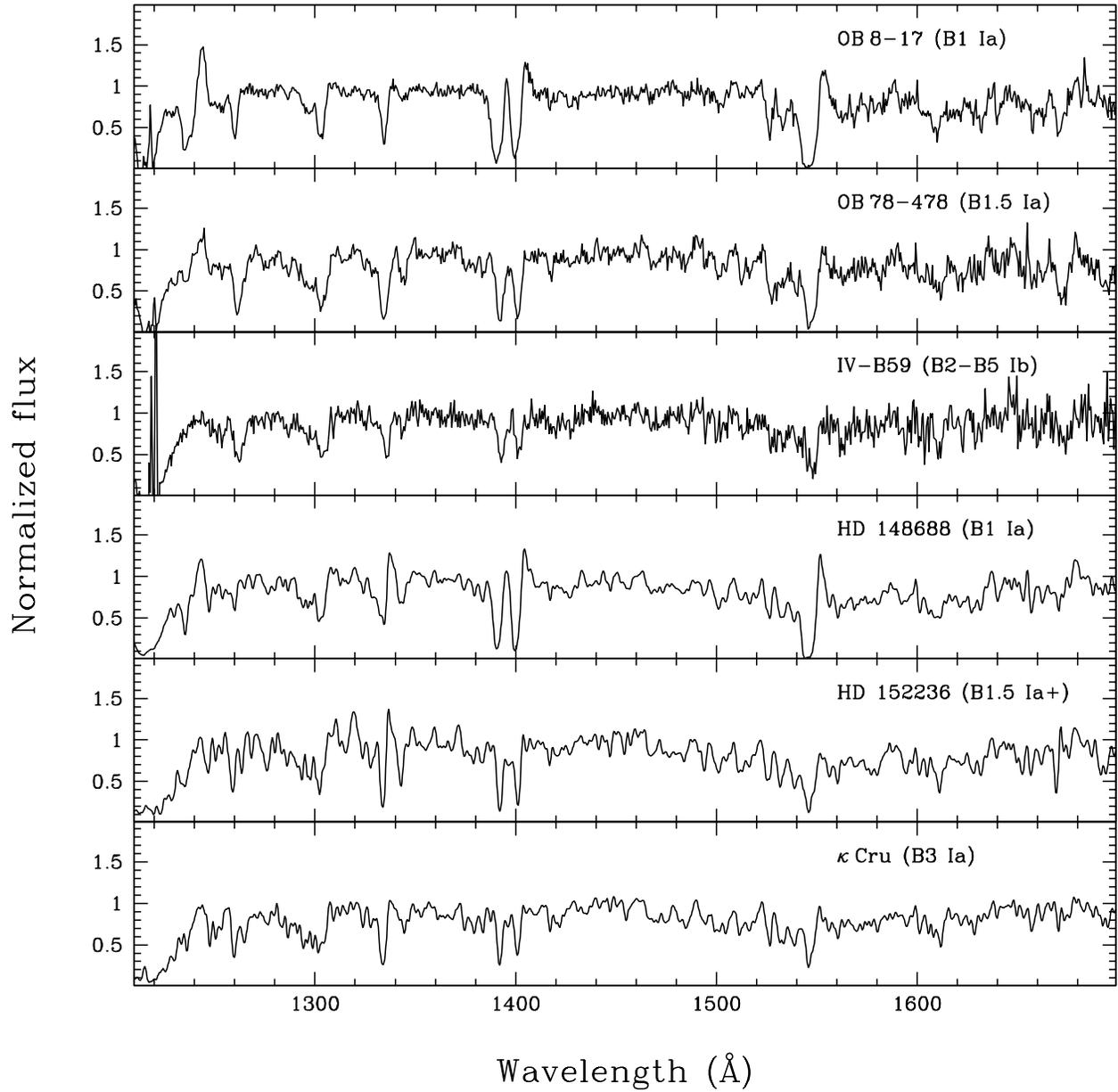}
\caption{The three later-type supergiants in the M31 sample 
(upper three panels) are compared to IUE spectra of Galactic stars of
similar spectral type, degraded to the resolution of the STIS spectra:
HD~148688 (B1\,Ia), HD~152236 (B1.5\,Ia+), and
$\kappa$~Cru (B3\,Ia) (lower three panels).\label{comp2}}
\end{figure}

In order to compare the UV spectral properties of our sample of M31
stars with counterparts observed in the Galaxy, we show in
Fig.~\ref{comp1} and \ref{comp2} our rectified STIS spectra, together
with selected Galactic OB supergiant spectra from the atlases of
\citet{walborn85} and \citet{walborn95}. The three top panels of
Fig.~\ref{comp1} illustrate the three B0\,Ia stars in M31, while the
remaining panels in this figure show IUE spectra of Galactic
supergiants ranging in spectral type from O9.7 to B0.5, degraded to
1.5~\AA\/ resolution. In a similar fashion, in Fig.~\ref{comp2} the
three later type supergiants in M31 are compared to B1-B3\,Ia Galactic
stars.  In the following we discuss the comparison on a star-by-star
basis, and briefly comment on the results obtained from the wind
analysis of the individual M31 supergiants.

\medskip

{\em OB\,78-159:} Low resolution spectra of this star, as well as of
OB\,8-17, OB\,10-64 and OB\,48-358, obtained with HST/FOS, were described
qualitatively by \citet{bianchi96}, who pointed out the general
similarity between the UV spectra of the M31 supergiants and Galactic
counterparts of the same optical spectral type.  The low spectral
resolution of their data, however, did not allow accurate quantitative
measurements of the terminal velocities.  The STIS acquisition images
show two fainter neighboring stars, the closest being $\sim$\,0.3
arcsec from the blue supergiant along the spatial direction in the
spectra, and which does not interfere with the extracted spectrum of
OB\,78-159. However, in future studies the luminosity inferred from
the ground-based apparent magnitude would have to be corrected using
high spatial resolution photometry with HST. The UV spectrum more
closely resembles that of the Galactic B0.5\,Ia template
($\kappa$~Ori) rather than the B0\,Ia template ($\epsilon$~Ori). It is
quite possible that the optical classification is compromised by the
contamination from the neighboring stars.  The measured value
$v_\infty=1200$~km\,s$^{-1}$ for the terminal velocity is also below
the average for Galactic B0 supergiants ($1550$~km\,s$^{-1}$),
although still within the range for B0\,Ia stars (\citealt{prinja90},
\citealt{howarth97}), and more in agreement with the average
$v_\infty=1380$~km\,s$^{-1}$ found for Galactic B0.5\,Ia stars
(\citealt{haser95}).

\medskip

{\em OB\,10-64:} A partial analysis of our STIS spectrum was included
in the work of \citet{smartt01}. We have slightly revised the terminal
velocity downwards given there from $v_\infty=1650$ km\,s$^{-1}$ to
1600 km\,s$^{-1}$, in order to obtain an agreement between the three
lines for which we attempted a profile fit, while the earlier result
was more weighted to the C\IV\w1550 line. This is the only supergiant
for which we currently have a mass-loss determination,
$\dot{M}=1.6\times10^{-6}$~\msun\,yr$^{-1}$, from the analysis of the
H$\alpha$ line (\citealt{smartt01}).  The Si\IV\/ wind line profile in
this star is more typical of O9-O9.7\,Iab supergiants rather than
B0\,Ia, as can be seen in Fig.~\ref{comp1}. The same can be said for
the spectrum of OB\,48-358. Within the uncertainties these two stars
have the same terminal velocity, which is considerably larger
($\sim$\,35\%) than the one measured in the other B0\,Ia star,
OB\,78-159.

\medskip

{\em OB\,48-358:} See comments for OB\,10-64.

\medskip

{\em OB\,8-17:} The star is classified as B1\,Ia by
\citet{trundle02}, and its UV spectrum well matches that of B0.7\,Ia or B1\,Ia
Galactic supergiants. The wind line N\V\w1240 in general becomes
weaker than observed here in Galactic supergiants later than B1\,Ia, and
also the other resonance lines in the spectrum of OB\,8-17 show a good
match with HD~148688 (B1\,Ia), as illustrated in
Fig.~\ref{comp2}. Indications for a match with the optical
classification comes also from C\III\w1175, which is blueshifted.  The
terminal velocity well agrees with that of typical B1\,Ia supergiants
in the Galaxy.

\medskip

{\em OB\,78-478:} \citet{massey85} first discussed IUE spectra of this
star in the OB association NGC~206, later reanalyzed by
\citet{bianchi91}. The appearance of the STIS spectrum is in agreement
with Galactic UV spectra corresponding to its optical spectral type
(B1.5\,Ia). The terminal velocity $v_\infty=550$~km\,s$^{-1}$ also
agrees with the typical value found for Galactic B1.5 supergiants.

\medskip

{\em IV-B59:} This is the coolest star in the sample, with a UV
spectrum consistent with the spectra of Galactic supergiants of type
B2.5-B4 Ia-Ib.  Earlier or later types are ruled out by the strength
of both the C\IV\/ and Si\IV\/ lines. This is in agreement with the
range in spectral type found from the optical
spectra. \citet{humphreys79} classified this star as B2\,Iae (with
H$\beta$ and H$\gamma$ filled in), while more recent WHT spectra
suggest a range between B2\,Ib and B5\,Iab (\citealt{smartt02}).  Wind
effects on the line profiles are rather weak, and an absorption
component on the blue side of the line profiles increases the
uncertainty in $v_\infty$.  The model fit to the Si\IV\/ doublet shown
in Fig.~\ref{fit2} was calculated without photospheric contribution,
because of the likely metallicity mismatch with the Galactic
template. Our result $v_\infty=630$~km\,s$^{-1}$ is in agreement
with the study of Galactic supergiants by
\citet{haser95}, who found average terminal velocities between 750 and
570 km\,s$^{-1}$ in the B2-B3 spectral type range. Individual values
between 405 and 830 km\,s$^{-1}$ were found at B3 by \citet{prinja90}.

\section{Discussion and conclusions}

In our previous work on the UV spectra of B supergiants in the
Galactic Cyg OB2 association (\citealt{herrero01}) we were able to
use some of the stellar parameters derived from optical spectra,
together with the known Galactic WLR (\citealt{puls96}) and an assumed
set of abundances, to estimate mass-loss rates and mean ionization
fractions.  Given the uncertainties in abundances and ionization
fractions for the current sample of B supergiants in M31, we defer a
more complete analysis to a future paper, where we will analyze WHT
and Keck optical spectra in order to measure the stellar parameters,
and in addition the mass-loss rates from fits to the H$\alpha$ line
profiles. When combined with the terminal velocities obtained in this
work, we will be able to derive a WLR for early B supergiants in
M31. In the following we will make some qualitative comparisons with
Galactic supergiants, and draw some preliminary conclusions.

The abundance gradient in M31 has been characterized by relatively few
studies of \hii\/ region emission lines (e.g.,~\citealt{dennefeld81};
\citealt{blair82}; \citealt{galarza99}). The large angular
extent of the galaxy and the general faintness of the \hii\/ regions
has precluded more extensive and complete spectroscopic surveys of the
gaseous abundances. Moreover, the low excitation of the nebulae
implies that the auroral lines necessary to determine the electron
temperature remain undetected, with the consequence that empirical
strong line methods must be used to estimate the abundances. A recent
reanalysis of the available \hii\/ region abundances, based on the
empirical abundance calibration of \citet{mcgaugh91}, has been
presented by \citet{smartt01}. The resulting oxygen abundance gradient
would imply a range between $12+\log (O/H)=8.9$ (at the radial
distance of OB\,8-17 and OB\,10-64) and 8.6 (IV-B59) for the current
sample of B-type supergiants, i.e., between a roughly solar abundance
and a 0.3 dex lower value.  On the other hand, the scant data
available from direct stellar abundance determinations
(\citealt{smartt01},
\citealt{venn00}) is consistent with a flat gradient between 5 and 20
Kpc, with a value close to that found in the Galactic solar
neighborhood ($12+\log (O/H)=8.7\pm0.1$, from the mean NLTE value of
\citealt{rolleston00}). The two methods (nebular vs.~stellar) of 
abundance determination might not be in disagreement, given the large
scatter in the derived nebular $O/H$ ratio at a given radial position,
and the small number of stellar data available. However, the result
obtained for OB\,10-64 by
\citet[$12+\log (O/H)=8.7$]{smartt01} is well below the value obtained
from \hii\/ regions at a similar galactocentric distance.

\begin{figure}
\plotone{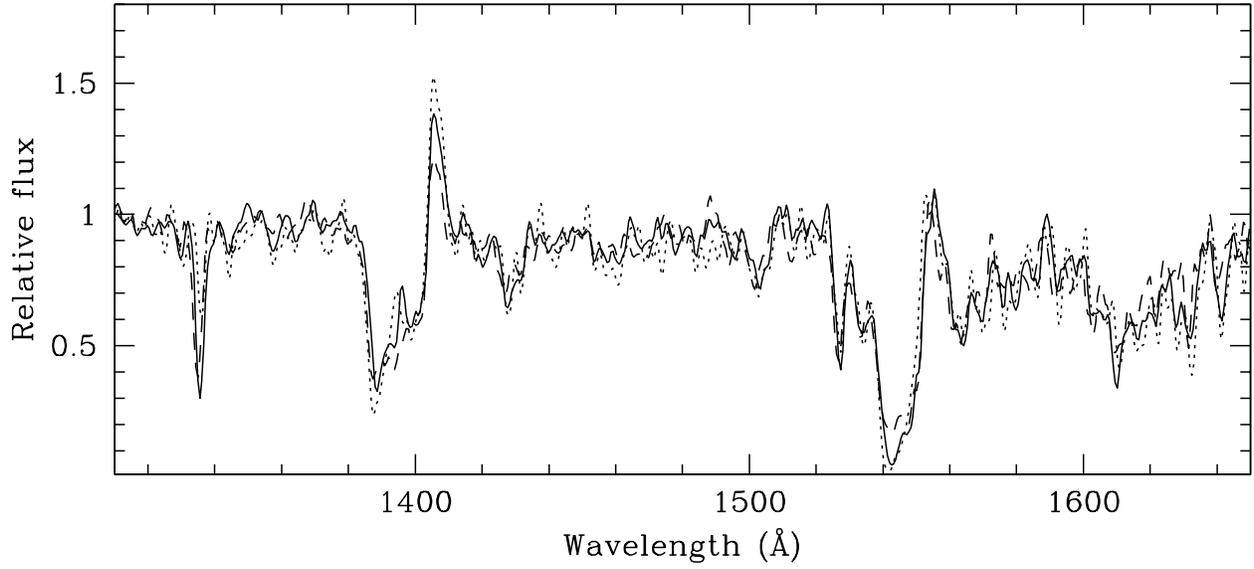}
\caption{The spectra of the two M31 B0\,Ia supergiants OB\,10-64 (full
line) and OB\,48-358 (dashed line) superimposed on the IUE
spectrum of HD~149038 (O9.7\,Iab, dotted line).\label{comp3}}
\end{figure}

\begin{figure}
\plotone{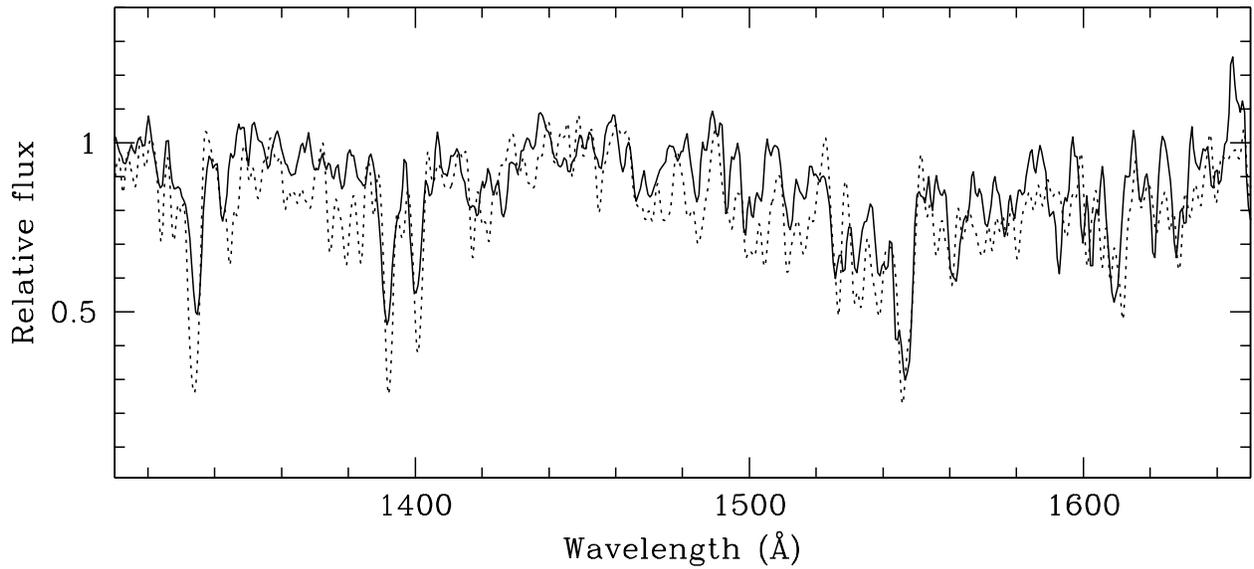}
\caption{The spectrum of IV-B59 (full line) superimposed on the IUE
spectrum of $\kappa$~Cru (dotted line).\label{comp5}}
\end{figure}

A detailed abundance study of the B supergiants observed with HST is
beyond the scope of the present paper, and is addressed in the
companion paper by \citet{trundle02}. Therefore here we directly
compare the UV spectra of B-type supergiants in M31 with stars of
similar spectral type in the Galaxy.  Any large discrepancies would
immediately reveal differences in the stellar abundances and/or the
wind properties. In Fig.~\ref{comp3} the spectra of two B0\,Ia
supergiants (OB\,10-64 and OB\,48-358) are shown superimposed on the
IUE spectrum of the Galactic O9.7 Iab star HD~149038. The latter was
retrieved from the IUE archive at STScI and rectified in the same way
as done for the M31 stars.  There is an overall excellent agreement
between the three spectra. In particular, we note a good match in the
spectral regions most affected by metal (mostly iron) lines, between
1410 and 1500~\AA, and at wavelengths redwards of C\IV\w1550. The
Si\IV\w1400 and C\IV\w1550 wind line profiles are also very similar in
the M31 and Galactic supergiants, with nearly the same terminal
velocities and peak intensities. A second example is illustrated in
Fig.~\ref{comp5}, where the UV spectra of IV-B59 and $\kappa$~Cru
(B3\,Ia) are compared. In this case a slight underabundance is
suggested in the spectrum of the M31 star, which would be in agreement
with its large galactocentric distance (22~Kpc). The WHT spectra of
IV-B59 are not of high enough quality for the accurate measurement of
metal lines, hence an abundance analysis of this star is not presented
in \citet{trundle02}. The \hii\/ region BA\,500 (\citealt{baadearp64})
lies only $10\arcsec$ east of IV-B59, and its emission line fluxes were
measured by \citet{dennefeld81}. Applying to their measurements the
analytical expression given by
\citet{kobulnicky99} for the determination of empirical abundances,
based on the strength of the [O\,II]\w3727 and [O\,III]\ww4959,5007
lines, and on the \citet{mcgaugh91} photoionization models, we obtain
$12+\log (O/H)=8.53$, which is approximately 2/3 of the oxygen
abundance in the solar neighborhood. However, the F5 supergiant A-207,
also located in Baade's Field IV, has been found to have a roughly
solar abundance by \citet{venn00}. More detailed abundance studies of
individual BA supergiants in M31 are clearly needed to clarify the
issue of the abundance gradient in this galaxy.

Similar qualitative comparisons have been shown by \citet[see also
\citealt{bianchi94}, \citealt{haser95b}]{bianchi96}, who compared the UV spectral
appearance of M31 and M33 early B supergiants with Galactic and LMC
ones. The effects of the lower metal abundance were clearly seen in
the M33 stars (see also \citealt{urbaneja02}), with weaker absorption
in the Si\IV\/ and C\IV\/ lines, and in the metal lines in general, in
a similar fashion to what is observed in the LMC stars. On the other
hand, in the case of M31 their results are comparable to those shown
in Fig.~\ref{comp3}-\ref{comp5}.

Our data (see Table~2), together with previous reliable determinations
of terminal velocities in M31 supergiants (\citealt{bianchi94};
\citealt{haser95b}; \citealt{bianchi96}), support the idea that the 
winds in M31 stars are comparable to those observed in the Galaxy.
This is also supported by the H$\alpha$ spectra studied by
\citet{herrero94} and by the differential analysis of the optical
spectra by \citet{trundle02}, which indicates that four of the stars
have very similar abundances to those derived in B-type supergiants in
our Galaxy within 1-2 Kpc of the Sun's position. The oxygen abundances
derived are listed for reference in column 3 of Table~2.  The absolute
value for OB\,8-17 appears significantly lower than the others at
8.4 dex.  However, a differential abundance analysis with two solar
neighbourhood supergiants indicates that OB\,8-17 has a very similar
oxygen abundance compared to these Milky Way stars. This is also
supported by a comparison of the UV spectrum of this object with that
of its Galactic counterpart, HD~148688.

\begin{figure}
\plotone{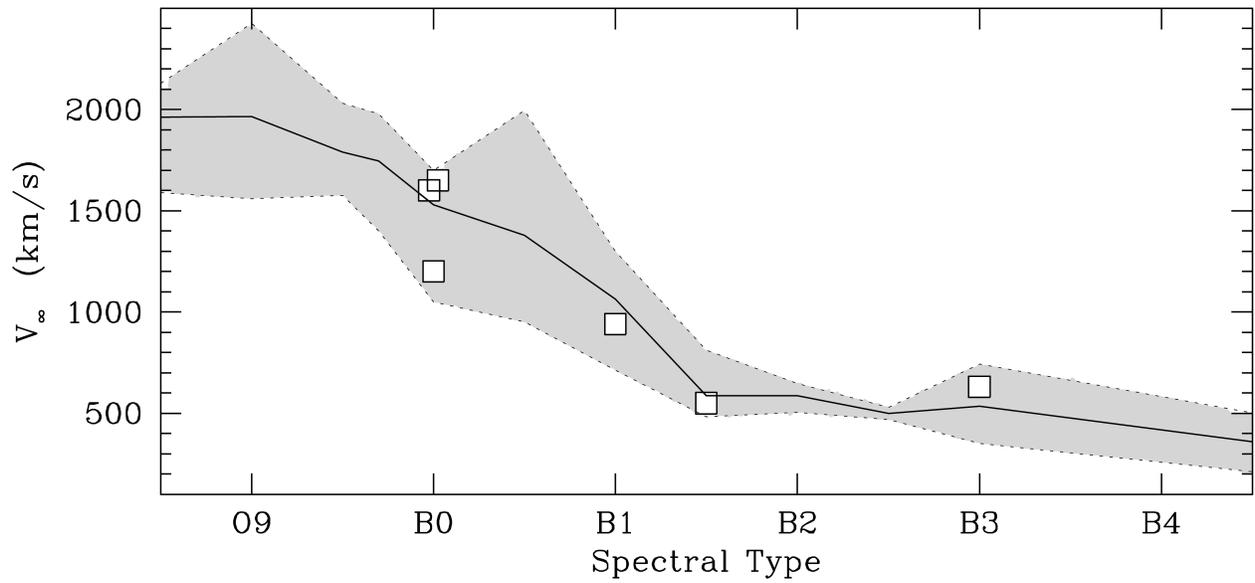}
\caption{The terminal velocities as a function of spectral type for the
M31 supergiants (open squares) are compared to the mean terminal
velocities from the compilations of \citet{haser95} and
\citet{howarth97} (full line). The shaded area shows the range in
terminal velocity for the Galactic stars. For star IV-B59 we have
assumed a B3 type, given the present uncertainty in its spectral
classification.
\label{vinf}}
\end{figure}

Although in one instance the measured $v_\infty$ is well below the
mean Galactic value for the given spectral type, all available data
lie within the range observed for Galactic stars, and are therefore
consistent with the expected scatter in M31. This is summarized in
Fig.~\ref{vinf}, where the Galactic data have been taken from the
compilations of terminal velocities of O and B supergiants by
\citet{haser95} and \citet{howarth97}. 
A similar conclusion was reached by \citet{bianchi96}, although their
terminal velocities are in general less accurate then the ones
measured with STIS, because of the lower resolution of some of their
data.

Our next step will be to add the information from the high-resolution
optical spectra, in order to derive wind momenta for the supergiants
studied here. This will allow us to provide better constraints on the
WLR for early B supergiants of solar abundance.

\acknowledgments
This work was supported in part by the German DLR under grant 50 OR
9909 2. AH and MAU thank the Spanish MCyT for financial support under
project AYA2001-0436.

%===================================================================================
%\clearpage

\clearpage
             
\begin{deluxetable}{lrrllllcrcl}
\tabletypesize{\scriptsize}
\tablecolumns{11}
\tablewidth{0pt}
\tablecaption{STIS targets in M31}
\tablehead{
\colhead{ID}	&
\colhead{R.A.}	&
\colhead{Decl.}	&
\colhead{$V$}		&
\colhead{$B-V$}		&
\multicolumn{2}{c}{Spectral Type} 	&
\colhead{$v_{rad}$} &
\colhead{Obs. date}	&
\colhead{	Exp. time}	&
\colhead{Comments}	\\
\colhead{}	&
\colhead{(2000.0)}	&
\colhead{(2000.0)}	&
\colhead{}	&
\colhead{}	&
\colhead{M95}	&
\colhead{T02}	&
\colhead{(km\,s$^{-1}$)}	&
\colhead{}	&
\colhead{(sec)}	&
\colhead{}}
\startdata
OB\,78-159  &  00 40 28.4  &  40 43 14.3 & 17.97 & $-0.05$ & B0 I & B0 Ia & $-467$ &
Feb 3, 1999 & 8223 & composite\\
OB\,10-64   &  00 44 10.4  &  41 33 16.2 & 18.10 & $-0.08$ & B1 I & B0 Ia & $-113$ &
Oct 1, 1999 & 8459 & Smartt et al. (2001)\\
OB\,48-358  &  00 45 13.2  &  41 39 38.4 & 18.70 & $-0.06$ & B0-1 I & B0 Ia & $-107$ &
Sep 29, 1999 & 14255 & \\
OB\,8-17    &  00 44 07.8  &  41 31 53.8 & 18.01 & $-0.06$ & O9-B1 I & B1 Ia & $-102$ &
Aug 9, 1999 & 8463 & \\
OB\,78-478  &  00 40 33.0  &  40 45 10.9 & 17.83 & $-0.10$ & B0-B1 Ib & B1.5 Ia & $-561$ &
Sep 26, 1999 & 5323 & \\
IV-B59\tablenotemark{a}      &  00 37 33.3  &  40 00 36.7 & 17.60 &
\phantom{$-$}0.00 & B2 Iae{\tablenotemark{b}} & B2 Ib-B5 Iab & $-502$ & Jul 7-8, 2000 & 5323 & \\
\enddata
\tablenotetext{a}{Baade \& Swope (1963)}
\tablenotetext{b}{Humphreys (1979)}
%\tablecomments{References for additional ID}
\end{deluxetable}

\begin{deluxetable}{lrccccclrcccc}
\tabletypesize{\scriptsize}
\tablecolumns{13}
\tablewidth{0pt}
\tablecaption{Derived parameters\label{results}}
\tablehead{
\colhead{ID}	&
\colhead{R}		&
\colhead{12+}		&
\colhead{$E(B-V)$\tablenotemark{b}}	&
\colhead{$M_V$}		&
\colhead{$\log N(HI)$}	&
\colhead{$N(HI)/E(B-V)_0$}	&
\colhead{$v_\infty$}	&
\colhead{$v_{ta}$}	&
\colhead{$\beta$}	&
\multicolumn{3}{c}{$\log N$ (cm$^{-2}$)}	\\
\colhead{}		&
\colhead{(Kpc)}		&
\colhead{$\log(O/H)$\tablenotemark{a}}	&
\colhead{}		&	
\colhead{}		&
\colhead{(cm$^{-2}$)}		&
\colhead{($10^{21}$cm$^{-2}$mag$^{-1}$)}		&
\colhead{(km\,s$^{-1}$)}	&
\colhead{(km\,s$^{-1}$)}	&
\colhead{}	&
\colhead{N\V}	&
\colhead{Si\IV}	&
\colhead{C\IV}	\\
\colhead{(1)}	&
\colhead{(2)}	&
\colhead{(3)}	&
\colhead{(4)}	&
\colhead{(5)}	&
\colhead{(6)}	&
\colhead{(7)}	&
\colhead{(8)}	&
\colhead{(9)}	&
\colhead{(10)}	&
\colhead{(11)}	&
\colhead{(12)}	&
\colhead{(13)}}
\startdata
OB\,78-159  & 9.5 & \nodata & 0.14	& $-6.93$ & 20.9 & \phantom{1}6.2 \phantom{1}(4.0) & 1200 \phantom{1}(50) & 150--350 & 0.8 & 15.24 & 15.34 & 15.70 \\
OB\,10-64   & 5.9 & 8.7 & 0.16	& $-6.86$ & 20.9 & \phantom{1}4.9 \phantom{1}(2.8) & 1600 (100) & 250--400 & 0.8 & 15.24 & 15.08\tablenotemark{c} & 16.52\\
OB\,48-358  & 11.6 & 8.7 & 0.23	& $-6.48$ & 21.0 & \phantom{1}4.1 \phantom{1}(1.8) & 1650 \phantom{1}(50) & 100--250  & 0.7 & 15.58 & 15.42 & 15.80 \\
OB\,8-17    & 5.8 & 8.4 & 0.16	& $-6.95$ & 20.9 & \phantom{1}4.9 \phantom{1}(2.8) & \phantom{1}940 \phantom{1}(50) & 200--300 & 0.7 & 15.59 & 15.59 & 15.83 \\
OB\,78-478  & 9.0 & 9.0 & 0.14	& $-7.07$ & 21.2 & 16.0 \phantom{1}(8.5) & \phantom{1}550 \phantom{1}(50) & 80--250 & 0.7 & \nodata & 15.42 & 16.64 \\
IV-B59 & 21.8 & \nodata & 0.14 & $-7.30$ & 21.2 & 16.0 \phantom{1}(8.5) & \phantom{1}630 (100) & 80--250 & 0.9 & \nodata & 14.84\tablenotemark{c} & 15.36 \\
\enddata
\tablenotetext{a}{\citet{trundle02}}
\tablenotetext{b}{\citet{massey95}; \citet{humphreys79}}
\tablenotetext{c}{No photospheric spectrum in fit}
\end{deluxetable}

\end{document}